\documentclass[aps,prl,amsmath,amssymb,preprint,superscriptaddress]{revtex4} 
\usepackage{graphicx}
\usepackage{dcolumn}
\usepackage{bm}

\begin{document}
\title{Modification of turbulent transport with continuous variation of flow shear in the Large Plasma Device}
\author{D.A. Schaffner}
\affiliation{Department of Physics and Astronomy, University of California, Los Angeles}
\author{T.A Carter}
\affiliation{Department of Physics and Astronomy, University of California, Los Angeles}
\author{G.D. Rossi}
\affiliation{Department of Physics, University of Texas, Austin}
\author{D.S. Guice}
\affiliation{Department of Physics and Astronomy, University of California, Los Angeles}
\author{J.E. Maggs}
\affiliation{Department of Physics and Astronomy, University of California, Los Angeles}
\author{S. Vincena}
\affiliation{Department of Physics and Astronomy, University of California, Los Angeles}
\author{B. Friedman}
\affiliation{Department of Physics and Astronomy, University of California, Los Angeles}
\date{\today}
\begin{abstract}
Continuous control over azimuthal flow and shear in the edge of the Large Plasma Device (LAPD) has been achieved using a biasable limiter which has allowed a careful study of the effect of flow shear on pressure-gradient-driven turbulence and transport in LAPD. LAPD rotates spontaneously in the ion diamagnetic direction (IDD); positive limiter bias first reduces, then minimizes (producing a near-zero shear state), and finally reverses the flow into the electron diamagnetic direction (EDD). Degradation of particle confinement is observed in the minimum shearing state and reduction in turbulent particle flux is observed with increasing shearing in both flow directions. Near-complete suppression of turbulent particle flux is observed for shearing rates comparable to the turbulent autocorrelation rate measured in the minimum shear state.  Turbulent flux suppression is dominated by amplitude reduction in low-frequency ($<10$kHz) density fluctuations. An increase in fluctuations for the highest shearing states is observed with the emergence of a coherent mode which does not lead to net particle transport. The variations of density fluctuations are fit well with power-laws and compare favorably to simple models of shear suppression of transport.
\end{abstract}
\maketitle
While flow shear does provide a source of free energy for instability and turbulence, it can lead to stabilization of pressure-gradient-driven instabilities and a reduction of turbulent transport in magnetized plasmas~\cite{burrell97, terry00}.  The transport barrier in the high-confinement mode, or H-mode, of tokamak operation~\cite{wagner82} is attributed to the spontaneous development of an edge flow layer in which  strong shearing suppresses transport~\cite{burrell97, terry00}.  The direct connection between the H-mode edge flow layer and improved confinement was first established in experiments on the Continuous Current Tokamak (CCT) in which transport barriers were generated by directly driving edge flow using torque due to radial currents driven by biased electrodes~\cite{taylor89}.  Biasing has been used to produce improved confinement in a number of subsequent experiments including toroidal devices~\cite{weynants92,boedo00,silva06} and linear magnetized plasmas~\cite{sakai93,maggs07,carter09}. 

While ample evidence for transport reduction in the presence of sheared flow exists~\cite{burrell99, tynan09} and significant effort and progress has been made in developing a theoretical understanding of the interaction between sheared flow and turbulence, there are still a number of open questions that can be answered by
experiment.  In particular, the exact mechanism behind turbulence modification and transport suppression by shear is still subject to debate: theories present a number of mechanisms including radial decorrelation~\cite{biglari90}, nonlinear reduction of turbulent amplitude~\cite{kim04}, and modification of turbulent cross-phase~\cite{ware96}.  Evidence for all of these mechanisms exists in experimental data~\cite{tynan09}, but a comprehensive experimental dataset establishing in detail the parameter regimes where each mechanism is important has not been acquired.  In part, this is due to
the fact that most datasets on flow-turbulence interaction come from studies of spontaneously generated flow or in cases where precise external control over flow and flow shear is not possible.  A number of basic plasma experiments have utilized biasing techniques to drive
flow and flow shear to study flow driven instabilities (e.g. \cite{amatucci96,jass72}); however, experiments have not been done in which precise external control over flow shear has been achieved in higher-density plasmas with drift-wave turbulence to systematically study the changes in turbulence characteristics and transport.

In this letter, we report on the first experiments in which external control of flow is used to document the response of turbulence and transport to a continuous variation of flow shear, including a zero shear state and a reversal of the flow direction. Shearing rates ($\gamma_{s}= \partial V_{\theta}/\partial r$, where $V_{\theta} = E_r/B$) from zero to up to five times the turbulent autocorrelation rate measured at zero flow shear $(\tau_{ac}^{-1})$ are achieved. Turbulent particle flux is reduced with increasing shearing rate, regardless of the direction of the flow or sign of the flow shear, with significant reduction occuring for $\gamma_{s} \sim \tau_{ac}^{-1}$.  The observed reduction in particle flux is dominated by a decrease in low-frequency ($f < 10$kHz) density fluctuation amplitude. For low frequency fluctuations, the crossphase between density and azimuthal electric field fluctuations remain near zero for all shearing rates.  With higher shear ($\gamma_{s} > \tau_{ac}^{-1}$) we observe the emergence of a coherent mode localized spatially in the region of strong flow. Fluctuations from this mode appear to increase density fluctuations above 10kHz, but do not appear to contribute to particle flux.   

The Large Plasma Device \cite{gek91} (LAPD) is a 17m long, $\sim$60cm diameter cylindrical plasma produced by a barium-oxide coated nickel
cathode. In the experiments reported here, a plasma of density $\sim$$2 \times 10^{12}$ cm$^{-3}$ and peak temperature of ~8eV is
produced in a uniform solenoidal magnetic field of 1000G.  Measurements of electron density, electron temperature, and potential (both plasma
potential and floating potential) are made using Langmuir probes.  
Measurements of ion saturation current ($I_{\rm sat} \propto n_e \sqrt{T_e}$) and floating
potential ($V_f$) are taken with a 9-tip Langmuir probe (flush-mount
tantalum tips) while temperature and plasma potential are
determined using a swept Langmuir probe. $I_{\rm sat}$ fluctuations are taken as a proxy for density fluctuations for the measurements reported in this work. Density profiles are determined by scaling averaged $I_{\rm sat}$ profiles to line-averaged interferometer measurements of density.  Turbulent particle flux
$\Gamma \propto \left<\tilde{n}_e \tilde{E}_\theta\right>$ is
determined through correlating density fluctuations from one tip
of this probe with
azimuthal electric field fluctuations ($E_\theta$) derived from
floating potential fluctuations on two azimuthally separated tips.
Azimuthal $E\times B$ flow is computed
using the swept-probe-derived plasma potential.  Flows derived using
this technique compare very well to measurements using
Mach probes~\cite{maggs07} and flows derived from time-delay
estimation (TDE) of the velocity of turbulent structures~\cite{holland04}.
  
Biasing experiments have been previously conducted on LAPD in which
edge profile steepening and a reduction in turbulent flux was
observed~\cite{maggs07,carter09}. In these experiments, edge flow was
driven by biasing the vacuum chamber wall with respect to the
plasma source cathode.  Transport reduction occurred only for biases
above a threshold value.  Below the threshold, azimuthal flow was
localized near the biased wall and no flow or flow shear was driven in
the region where drift wave turbulence exists.  Above the threshold,
the flow was able to penetrate radially inward; hence, strong flow and
flow shear, with shearing rates far above the low-flow turbulent
autocorrelation rate, was driven in the region of strong density
gradient.   Recent experiments were successful in achieving more continuous control of potential and cross-field flow in the shadow of a small biased obstacle inserted into the LAPD core plasma~\cite{zhou12}.  Both confinement improvement and degradation (formation of strong density depletions) were observed with the density profile created by the obstacle in this case.  

\begin{figure}[!htbp]
\centerline{
\includegraphics[width=8.5cm]{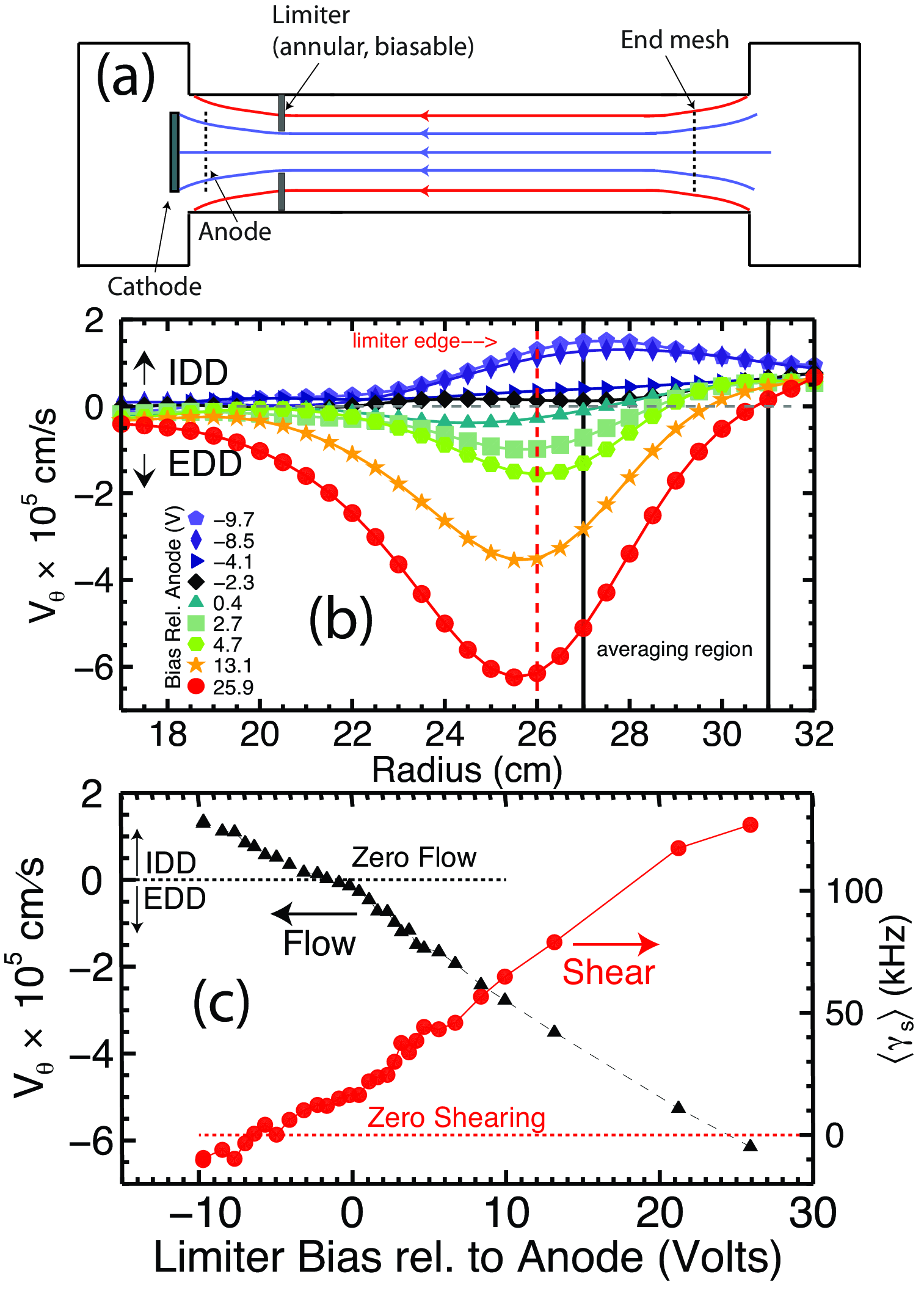}}
\caption{\label{fig:velocity_flowshear} (a) Diagram of the LAPD device showing annular limiter.  (b) Velocity profiles using plasma potential from swept measurements. (c) Flow at the limiter edge (black, triangles) and mean shearing rate, averaged over $27 < r < 31$cm (red, circles).}
\end{figure}

Motivated by the success of biasing obstacles to control flow, a large
annular aluminum limiter was installed in LAPD. The limiter provides a
parallel boundary condition for the edge plasma and is biased relative
to the cathode of the plasma source to control plasma potential and
cross-field flow.  The limiter is an iris-like design with four
radially movable plates located 2.5m from the cathode as shown
schematically in Fig.~\ref{fig:velocity_flowshear}(a).  The limiters
create a 52cm diameter aperture; downstream of the limiter, plasma on
field lines with radial location $r>26$cm has the limiter as a
conducting end parallel boundary condition and plasma on field lines
for $r<26$cm has the anode/cathode of the source region as a parallel
boundary condition.  An electrically floating conducting end mesh
terminates the plasma on the far end of the device.  A capacitor bank
and transistor switch supply a voltage pulse to the limiter.  The bias
pulse lasts 5ms during the flat-top of the $\sim$$15$ms plasma
discharge. The limiter is biased from $\sim$$10$V below to 50V above
the anode potential.  Typically, plasma potential in the core LAPD
plasma (plasma on field lines that connect to the source region) is
very close to the anode voltage and the cathode sits near ground
(vacuum chamber wall).  The anode potential is above the cathode
potential by the discharge voltage, which was $\sim$$40$V during these
experiments.

Spontaneous azimuthal rotation of the LAPD plasma is observed when the limiters are
unbiased (here the limiters are observed to float to a
potential $\sim 10$V below the anode).  In this state, an edge flow
(peaked just outside the limiter edge) is
observed in the ion diamagnetic drift direction (IDD), as shown in
Figure~\ref{fig:velocity_flowshear}(a).  Biasing the limiter positively
with respect to the cathode tends to drive flow in the electron
diamagnetic drift direction (EDD).  As the limiter bias is increased, the
flow in the IDD is first reduced, then brought to separate near-zero flow
and zero flow-shear states, and ultimately reversed with strong EDD flow.

Measurements of profiles of density and particle flux
were made for each bias flow state. Values are averaged over a range
from $r=27$cm to $r=31$cm, a region where average flow and flow shear scale
nearly linearly with limiter bias, as shown in
Figure~\ref{fig:velocity_flowshear}(b).  All other spatially-averaged
quantities shown in this paper are averaged over the same region in space.

\begin{figure}[!htbp]
\centerline{
\includegraphics[width=8.5cm]{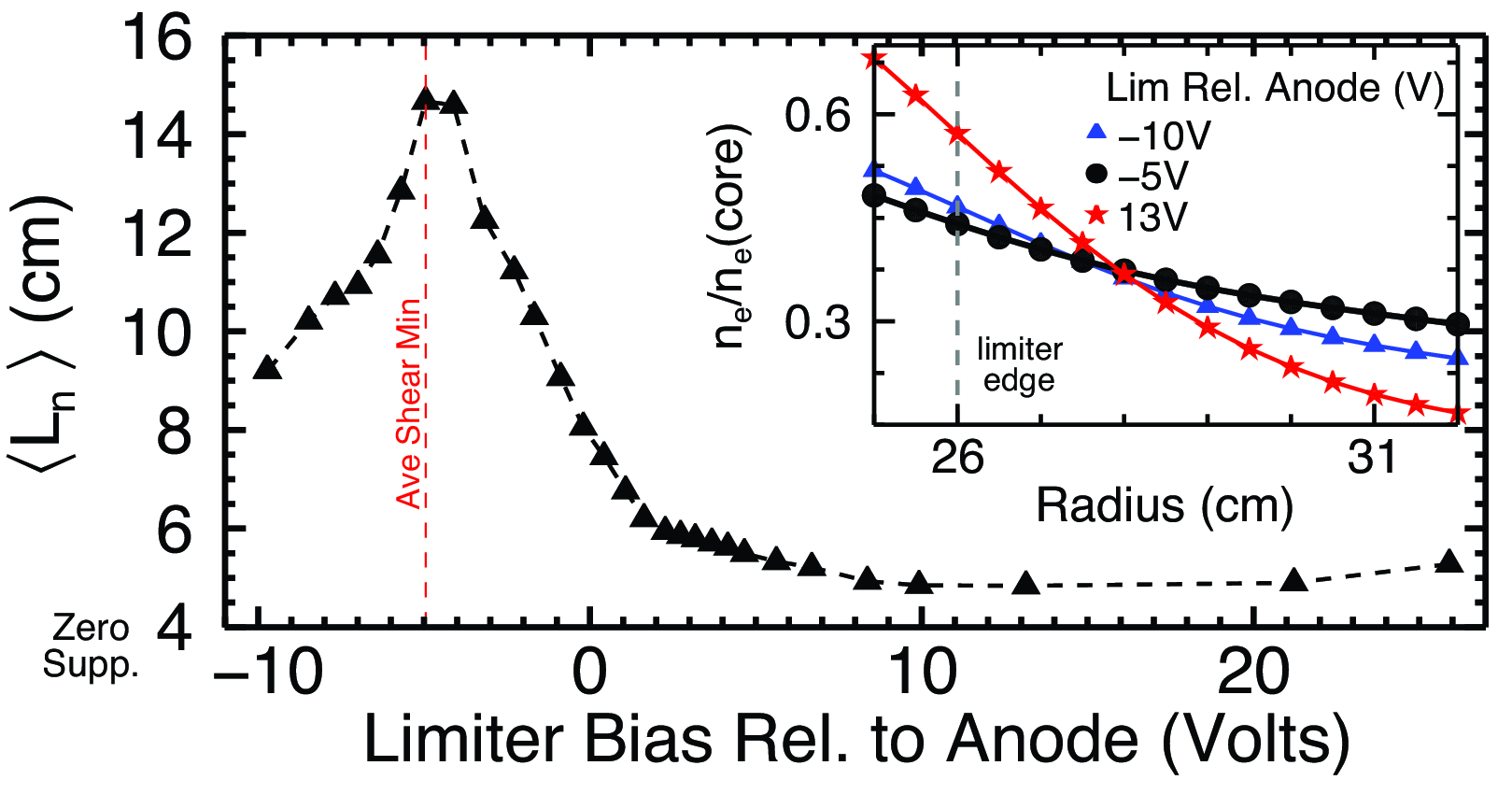}}
\caption{\label{fig:densgrad} Density gradient length scale versus limiter bias. Inset shows density profile at three bias values.}
\end{figure}

Figure~\ref{fig:densgrad} shows the variation in the spatially-averaged density gradient length scale, $L_{n} = \lvert \nabla \ln n \rvert ^{-1}$ with
increasing limiter bias.  As the limiter bias is increased, reducing
the IDD flow, an increase in the gradient scale length is observed,
indicating a degradation of radial particle confinement. The gradient scale length peaks
when the averaged shearing rate is near zero. As the bias is
increased further, reversing the flow and again increasing the
shearing rate, the gradient gradually steepens and the
scale length is lowered, indicating improved radial particle confinement.  

\begin{figure}[!htbp]
\centerline{
\includegraphics[width=8.5cm]{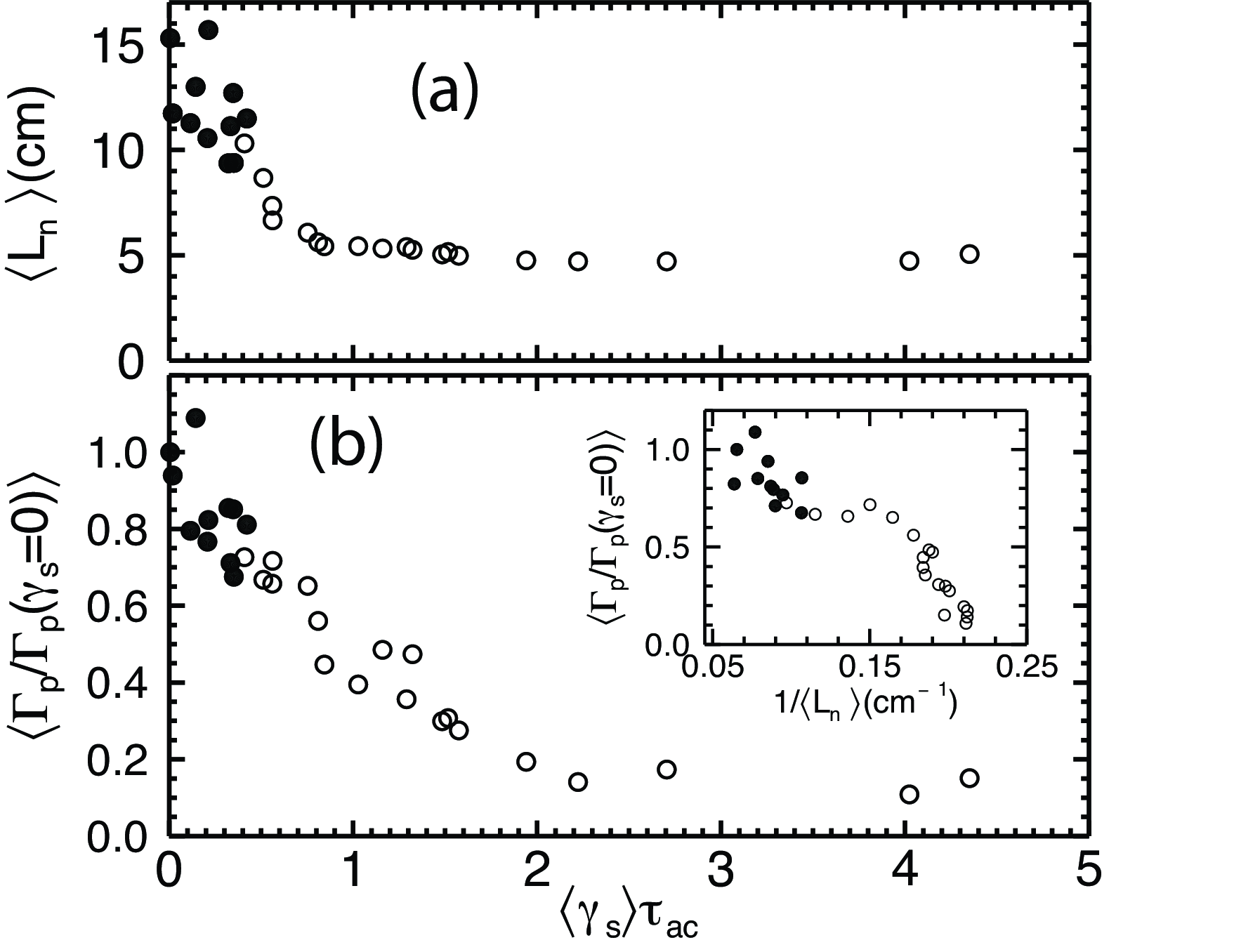}}
\caption{\label{fig:sheargradflux} (a)Gradient scale length versus shearing rate. (b)Particle flux normalized to no-shear
  flux as a function of normalized shearing rate. Filled symbols
  represent points with flow in the IDD. Inset: Measured turbulent particle flux versus
  gradient scale length.}
\end{figure}

The observed variation of $\langle L_{n} \rangle$ with bias is best
organized when compared to the shearing rate, $\gamma_{s}$, as is
shown in Figure~\ref{fig:sheargradflux}(a).   The shearing rate is
normalized to the autocorrelation rate of density fluctuations
measured in the zero-shear state.  An autocorrelation rate of $\tau_{ac}^{-1} \approx $ 28kHz $(\tau_{ac} \approx 36\mu s)$ is calculated by taking the half-width at half-maximum of a Hilbert transform of the $I_{\rm sat}$
autocorrelation function.  Confinement improvement (decreased $\langle
L_n \rangle$) occurs continuously and gradually with increasing
$\gamma_{s}$ and reaches saturation for $\gamma_{s} \approx \tau_{ac}^{-1}$ (a normalized $\gamma_{s}$ of 1).  The profile steepening
appears to be largely independent of the direction of the flow (or radial electric field): IDD (filled points) and EDD (open points) flow cases follow the same trend when plotted against normalized shearing rate.

Measured changes in turbulence and turbulent particle flux are
consistent with the observed changes in the density profile.  The
turbulent particle flux can be written\cite{powers74}:
\begin{equation}
\Gamma = \frac{2}{B} \int^{\infty}_{0} \lvert n(f) \rvert \lvert E_{\theta}(f) \rvert \gamma_{(n,E_{\theta})}(f) \cos [\phi_{(n,E_{\theta})}(f)] df
\label{eq:fluxint}
\end{equation}
where $n(f)$ and $E_\theta(f)$ are the Fourier transforms of
the density and azimuthal electric field fluctuations;
$\gamma_{(n,E_\theta)}$ is the coherency between density and electric
field; and $\phi_{(n,E_\theta)}$ is the cross-phase angle between
density and electric field.

Figure~\ref{fig:sheargradflux}(b) shows the spatially-averaged turbulent
particle flux as a function of normalized shearing rate.  The
turbulent flux decreases continuously with increasing shearing rate;
however the observed decrease is slightly slower than that observed
for $L_n$.  The inset in Figure~\ref{fig:sheargradflux}(b) shows that the variation in
turbulent flux is correlated with the changes in $L_n$ (but scales in a way
that is inconsistent with Fick's law using a fixed diffusion coefficient).  The
trend in reduced particle flux is the same for either direction of
flow (IDD or EDD).  The cause for the reduction in turbulent particle
flux can be explored by considering individual terms in the integrand
of Eqn.~\ref{eq:fluxint}.

\begin{figure}[!htbp]
\centerline{
\includegraphics[width=8.5cm]{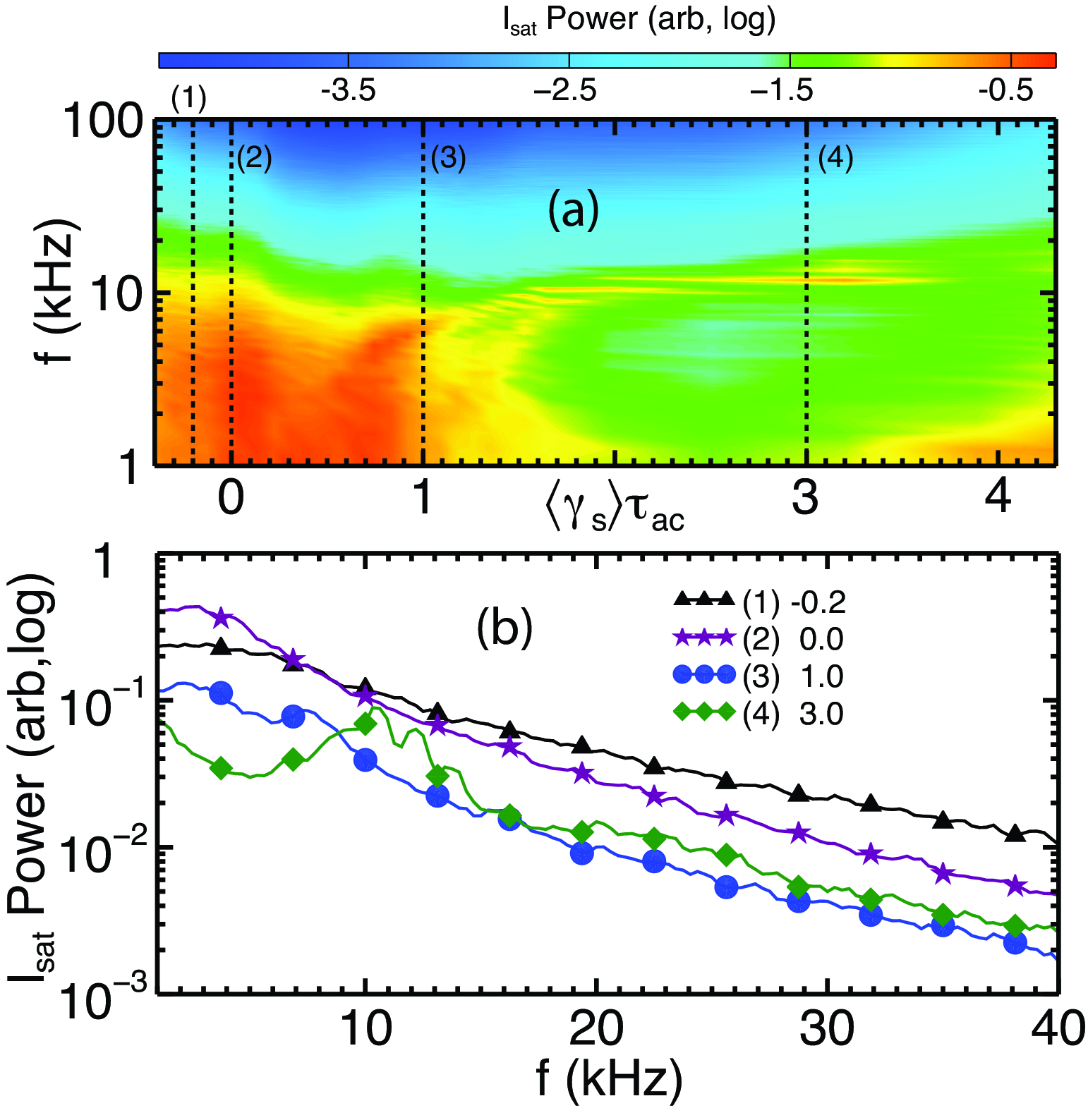}}
\caption{\label{fig:powercontour} (a) Contour plot of log $I_{\rm sat}$/density fluctuation power versus shearing rate and frequency. (b) Power spectra for four different values of shearing rate.}
\end{figure}

Density fluctuations were reduced significantly with increasing
shearing in these experiments.  Figure~\ref{fig:powercontour}(a) shows
changes in the spatially-averaged density fluctuation spectrum
with shearing rate.  The shearing rate is signed in this figure, and
negative shearing rates occur for flow in the IDD. Most of the power
is located in frequencies $<10$kHz and in this range, power decreases
overall with increasing shearing rate.  A decrease of about one order
of magnitude in fluctuation power is seen between the minimum shear
state and the high shear regime where $L_n$ and particle flux are
minimized; this is made clearer in Figure~\ref{fig:powercontour}(b).  At
higher shearing rates, $\gamma_{s} \gtrsim \tau_{ac}^{-1}$, a coherent
mode emerges.  The frequency of the mode increases with shearing rate
and the fluctuation amplitude is localized to the peak of the
azimuthal flow.

\begin{figure}[!htbp]
\centerline{
\includegraphics[width=8.5cm]{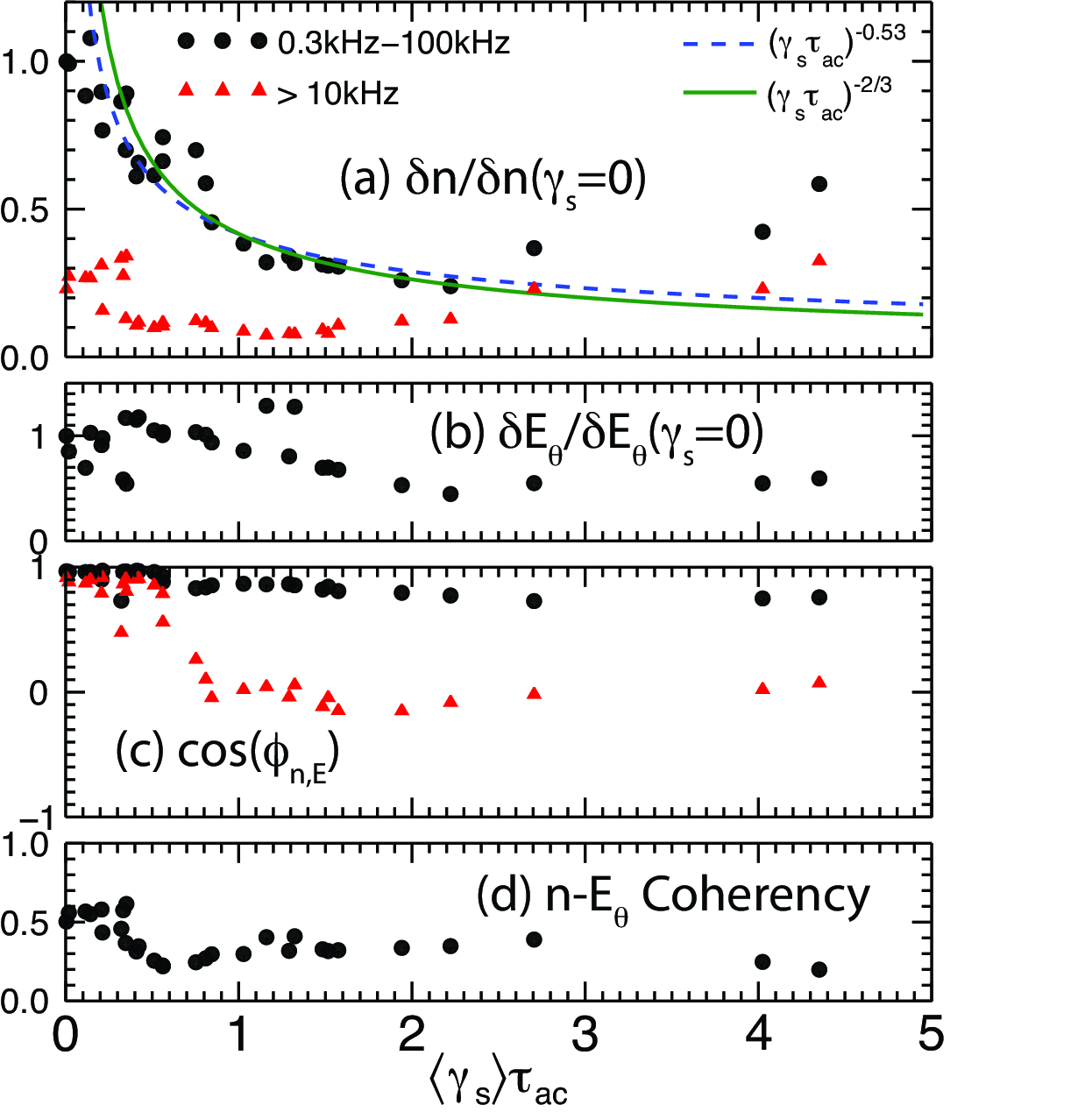}}
\caption{\label{fig:fluxcomps} Components of particle flux versus shearing rate including $I_{\rm sat}$/density fluctuation power(a), electric field fluctuation power(b), crossphase(c) and coherency(d) with black points for low or all frequency, red for high only.}
\end{figure}

Figure~\ref{fig:fluxcomps}(a) shows the reduction in total density fluctuation amplitude with shear in two frequency bands: all frequencies below 100kHz in black and all frequencies above 10kHz in red. With the emergence of the coherent mode, the
high frequency fluctuation amplitude does show an increasing trend at higher shearing
rates but there is a strong overall decrease in fluctuation amplitude with shearing.
A reduction is also seen in $E_\theta$ fluctuation amplitude, as shown in
Figure~\ref{fig:fluxcomps}(b); however this reduction is weaker than
observed in density fluctuations.  The cross-phase between
$n$ and $E_\theta$ does not change significantly with
shearing. As shown in Figure~\ref{fig:fluxcomps},
$\cos[\phi_{(n,E_\theta)}] \sim 1$ for all shearing rates.  For
higher frequencies ($f > 10$kHz), the cross-phase does change with
shearing, with $\cos[\phi_{(n,E_\theta)}]$ trending toward zero at
higher shear.  This crossphase change explains why the coherent mode
that emerges at higher shearing rate does not contribute to an
increase in the particle flux.  The coherency between $n$ and
$E_\theta$ also decreases with shearing rate, as shown in 
Figure~\ref{fig:fluxcomps}.  Overall, the decrease in flux is primarily
due to a decrease in turbulent amplitude.  This observation is distinct from previous work with flows driven by vacuum-chamber-wall
biasing on LAPD. In those experiments, turbulent amplitude decreased little while the
turbulent cross-phase experienced a significant change, leading to
reduced particle flux~\cite{carter09}.  In the experiments reported
here, the magnetic field is higher (1000G versus 400G) and normalized
shearing rates are lower (near unity).  Cross-phase change is expected
in cases with very strong shearing ($\gamma_{s} \gg \tau_{ac}^{-1}$)~\cite{terry01}.  Future experiments will explore the
variation of the turbulent response to higher normalized shearing
through changing plasma parameters, in particular magnetic field.  

Lastly, we add a comparison of our data to a simple theory, the
Biglari-Diamond-Terry (BDT) model~\cite{biglari90}, which predicts a
power-law scaling with shearing rate of the turbulent
amplitude of the form: $\left(\gamma_{s}/\tau_{ac}^{-1}\right)^{-\alpha}$. As seen in
Figure~\ref{fig:fluxcomps}, a best fit of $\alpha = 0.530$ compares
favorably to the BDT prediction of $\alpha = 2/3$ for the reduction in
density fluctuation amplitude. 
It should be noted, however, that the
BDT model is fairly simple and the validity of its assumptions is
questionable for the experimental conditions reported here.  In
particular, as the shearing rate is increased in LAPD, the density
profile is changing (in BDT a fixed drive is considered).  Future work
will focus on direct comparisons to more comprehensive models of shear
suppression, including comparisons to two-fluid simulations using the
BOUT++ 3D turbulence code~\cite{umansky11}.  

This letter presents the first experiments in which the response of
pressure-gradient-driven turbulence to a continuous
variation of shearing rate, including a near-zero flow shear state and
a reversal in the direction of flow, is studied.  Increased shearing
improves radial particle confinement regardless of the direction of
the azimuthal flow or sign of the flow shear. The observed reduction of
turbulent particle flux with shear is attributed to a reduction in the
amplitude of density fluctuations. These
experiments were performed at a fixed set of plasma parameters (fixed
magnetic field, neutral pressure, discharge power); future work will
explore the variation in turbulent response to shear as these
parameters are varied.  

\providecommand{\noopsort}[1]{}\providecommand{\singleletter}[1]{#1}%

\end{document}